**Pauli spin blockade in cotunneling transport through a double quantum dot**

H. W. Liu,[1,2,3] T. Fujisawa,[1,4] T. Hayashi,[1] and Y. Hirayama [1,2]

[1]NTT Basic Research Laboratories, NTT Corporation, 3-1 Morinosato-Wakamiya, Atsugi, 243-0198, Japan

[2]SORST-JST, 4-1-8 Honmachi, Kawaguchi, Saitama 331-0012, Japan

[3]National Laboratory of Superhard Materials, Institute of Atomic and Molecular Physics, Jilin University, Changchun 130012, P.R. China

[4]Tokyo Institute of Technology, 2-12-1 Ookayama, Meguro-ku, Tokyo 152-8551, Japan

We study spin correlation in a double quantum dot containing a few electrons in each dot ($\lesssim$ 10). Clear current rectification with negative differential conductance is observed in the cotunneling regime, which is well explained by a Pauli spin blockade. The current feature is discussed in connection with exchange singlet-triplet splitting based on two simple models, one of which takes coherent interdot coupling and the other incoherent interdot coupling into account.

The two-electron system in a double quantum dot (DQD) is attractive for studying fundamental spin correlations. The exchange Coulomb interaction between the electrons induces a singlet-triplet (ST) splitting, which has been discussed as a way to demonstrate spin logic gates (NOT, SWAP, etc.) and entangled states.[1-3] Spin correlation also leads to a Pauli spin blockade, at which the transport is blocked when the two electrons have spin triplet correlation.[4] This kind of spin-dependent tunneling has possible application to spin-qubit readout.[5] More recently, a Pauli spin blockade with nearly degenerated ST splitting has been employed for studying hyperfine coupling between electron and nuclear spins.[6,7] Besides appearing in the two-electron system, these spin effects may appear generally in a DQD with more than two electrons. In this article, we investigate Pauli spin blockade and ST splitting in an *effective* two-electron DQD containing about 10 electrons in each dot. Pauli spin blockade transport is clearly identified by the current rectification with negative differential conductance (NDC) in the cotunneling regime, where the ST splitting can be detected with improved resolution.[8] We discuss the cotunneling transport between the *effective* two-electron states by considering the coherent and incoherent coupling between the dots, which will be helpful for studying voltage-controlled exchange splitting in a DQD.

The left (L) and right (R) dots in our sample are electrostatically defined in an AlGaAs/GaAs heterostructure (inset of Fig. 1).[9] We control the tunneling rate $\gamma_L(\gamma_R)$ of the left (right) barrier separating the left (right) dot and the source (drain) electrode and the interdot coupling $t_c$ of the central barrier by voltages applied to gates $G_{L(R)}$ and $G_C$, respectively. All experiments were performed using dc current measurement at zero magnetic field (unless noted otherwise) at an electron temperature of $T_e \sim 25$ mK (similar to the lattice temperature at ~ 25 mK). Fig. 1 shows a color plot of current $I$ flowing through the double dot as a function of gate voltages $V_{GL}$ and $V_{GR}$, which change the equilibrium electron numbers $(n,m)$ in the left and right dots, respectively. Each dot might have accommodated a few electrons ($\lesssim 10$), but the labeled *effective* electron numbers $n$ and $m$ in the range from 0 to 2 explain our experiments well. Sequential tunneling current flowing through the dots appears at the corners of the hexagons, where three charge states are energetically degenerated.[10] Current is also visible along some sides of the hexagons, where one charge state is energetically higher than the other two. Cotunneling along the shorter sides (upward-right direction) conserves the total electron numbers ( $N = n + m$ ) in the two dots (charge-conserved process), while cotunneling along the longer sides changes N by 1 (charge-unconserved process). The current peaks in regions I and II involve charge-conserved cotunneling and sequential tunneling, while the charge-unconserved cotunneling is well suppressed. In these regions, we can effectively consider tunneling transitions between the N-electron states only.

When a finite source-drain voltage $V_{SD}$ is applied, the conductive region spreads out in the $V_{GL}$-$V_{GR}$ plane as shown schematically in Fig. 2(a). Sequential tunneling and charge-conserved cotunneling regions are indicated by hatched and dotted patterns, respectively. For convenience, we introduce two sweeping voltages, $V_\delta$ and $V_\varepsilon$. Changing $V_\delta$ shifts the average electrochemical potential $\delta$ of the two charge states ($n$,$m$) and ($n$-$1$,$m$+$1$), and changing $V_\varepsilon$ gives the energy difference $\varepsilon$ between the two states. The origin O ($V_\delta = V_\varepsilon = 0$, $\delta = \varepsilon = 0$) is taken at the resonant condition in the middle of the N-electron Coulomb blockade region. This diagram can be compared to the magnified plots in region I at positive and negative $V_{SD}$ which are shown in Figs. 2(b) and (c) respectively. No clear sequential tunneling current is resolved, while current in the cotunneling regime is clearly observed. This reminds us of the orbital Kondo effect, in which higher-order tunneling processes screen the orbital (charge) degree of freedom.[11] Along the $V_\varepsilon$ direction, the clear current peak at $\varepsilon = 0$ indicates resonant tunneling between the dots. The peak has a shoulder on the positive (negative) $V_\varepsilon$ side at positive (negative) $V_{SD}$ due to weak inelastic interdot tunneling with phonon emission.[12] No obvious bias-polarity dependence is observed, indicating there is no spin blockade in region I. When $V_{SD} \geq 1.1$ mV, another resonant line related to the (0,1) or (1,0) excited states appears (data not shown). We find that our dot contains more than one electron because the excitation spectrum of dot R with “m = 0” shows an excited line 0.4 meV above the ground state. Thus, the electron transport in region I denotes the tunneling transitions between *effective* one-electron (1,0) and (0,1) charge states.

Then neighboring region II or III in Fig. 1 may show the Pauli spin blockade. The blockade phenomenon is not clearly observed in region III, where relatively strong charge-unconserved cotunneling might have smeared out the blockade. However, obvious bias-polarity dependence is observed in region II, as shown in Figs. 2(d) and (e). In this case, sequential tunneling and charge-conserved cotunneling current can be identified clearly. Suppose the (0,2) ground state is a spin singlet. The negative-bias case corresponds to the tunneling from the (0,2) to the (1,1) singlet state. The transition in this direction proceeds freely and thus shows relatively large current [Fig. 2(e)]. At positive bias, however, once the system is excited to the (1,1) triplet state, an electron can not tunnel to the (0,2) singlet state due to Pauli exclusion and the blockade state is maintained until a spin-flip transition to the (1,1) singlet state occurs. Thus, current is strongly suppressed compared to the negative-bias case [Fig. 2(d)]. When $V_{SD}$ increases further, the spin blockade is lifted if the electron in the (1,1) triplet state is allowed to tunnel to the (0,2) triplet excited state with two parallel electron spins occupying different orbitals in dot R. The ST splitting of the (0,2) charge states is about 60 μeV and a magnetic field induces a singlet-triplet transition at 0.75 T (data not shown). These observations resemble the reported spin blockade in a true two-electron DQD.[4,13] The Pauli spin blockade is now clearly demonstrated even when the DQD contains more than two electrons.

Cotunneling current *I* as a function of $V_\varepsilon$ and $V_{SD}$ is shown in Fig. 3(a) at $\delta = 0$. At negative $V_{SD}$,

a clear resonant peak is observed around $\varepsilon = 0$ (dashed line), and inelastic interdot tunneling with phonon emission appears at $\varepsilon < 0$. This is qualitatively the same as the one-electron case in region I. At positive $V_{SD}$, the blockade is identified in the low-current region (~ 1 pA). High current (>> 5pA) in the upper-right region (ES) is associated with the transport through the (0,2) triplet state, where the blockade is lifted. We focus on the region of the small current peak with NDC at $V_{SD} = 5 \sim 20$ μV. Current at some constant $\varepsilon$ around the peak (parallel to the $\varepsilon = 0$ line) as a function of $V_{SD}$ is replotted in Fig. 3(b). In this way, we can compensate the electrostatic coupling between the dot and the electrode and discuss the true I-$V_{SD}$ characteristic at a given $\varepsilon$. A pronounced peak with NDC is observed in the range of $\varepsilon = -3 \sim 25$ μeV.

Figure 4 summarizes the zero-bias conductance $G_0$, the spin blockade current $I_{SB}$ at $V_{SD} = 50$ μV, and the current peak position $V_{peak}$ as a function of $\varepsilon$ and $\delta$. The two peaks (A and B) in the $\delta$-dependence of $G_0$ [Fig. 4(b)] correspond to sequential tunneling processes, from which we obtain the electrostatic coupling energy U (= 255 μeV). In order to avoid the influence of the sequential tunneling, the cotunneling region is confined to the $\delta$ range of -90 to 90 μeV (between the two solid straight lines), where the measured conductance is much larger than the sequential tunneling conductance shown by the dash-dotted Lorentzian curves. We fit $G_0$, $I_{SB}$ and $V_{peak}$ with two simple models, one of which takes coherent interdot coupling and the other incoherent interdot coupling into account: Note that cotunneling between the N = 2 charge states proceeds through the virtual intermediate (0,1) or (1,2) state with excitation energy $U/2 - \delta$ or $U/2 + \delta$. These virtual processes can be described in the cotunneling rate,[14,15] which simplifies our models and allows experimental access to resolve the ST splitting of (1,1) states.

In the first model, coherent coupling between the two dots is assumed and decoherence and the thermal effect are neglected. Coherent coupling between the (1,1) and (0,2) singlet states induces bonding (B) and antibonding (A) splitting [Fig. 3(c)]. Neglecting the spin-orbit effect, the (1,1) triplet state (T) is not affected by the coherent coupling. When $|eV_{SD}| < \Delta_B$ ( $\Delta_B$ is the energy spacing between the B and T states), elastic cotunneling through the B state is dominant and linear conductance is expected. When $eV_{SD}$ is larger than $\Delta_B$, inelastic cotunneling from the B to T states leads to the blockade. Therefore, a detailed analysis of the spin blockade enables us to extract $\Delta_B$. Possible cotunneling transitions between the three states are schematically drawn in the inset of Fig. 3(c). The elastic cotunneling rate $\Gamma_{B\pm}$ through the B state for positive (+) and negative (-) current contributes to the linear conductance:[14]

$$G_0 = e\frac{d(\Gamma_{B+} - \Gamma_{B-})}{dV_{SD}}\Big|_{V_{SD}=0} = \frac{e^2}{4h}\hbar^2\gamma_e^2 V^{-2}\sin^2\theta, \quad (1)$$

where $\theta = \tan^{-1}(\frac{2\hbar t_c}{\varepsilon})$ is the coupling angle ($0 \le \theta < \pi$) and $\sin^2\theta = 4(\hbar t_c)^2/[\varepsilon^2 + 4(\hbar t_c)^2]$. The effective excitation energy V is given by $V^{-1} = (U/2+\delta)^{-1} + (U/2-\delta)^{-1}$. The $\gamma_e$ is defined as the effective tunneling rate for elastic cotunneling involving higher-order tunneling processes (e.g., the Kondo effect), and can be simplified to $\gamma_e = \sqrt{\gamma_L\gamma_R}$ by considering only the second-order tunneling.

When large $V_{SD}$ is applied, current is given by a complicated function involving all possible cotunneling processes. However, in the spin blockade region, $I_{SB}$ is approximately determined by the lifetime of the (1,1) triplet state.[14] In this case, inelastic cotunneling from the T to B states with the tunneling rate $\Gamma_{tb}$ dominates the spin relaxation process. $I_{SB}$ is given by

$$I_{SB} = e\Gamma_{tb} = \frac{e}{h}\hbar^2\gamma_s^2 V^{-2}\Delta_B\sin^2\frac{\theta}{2}, \quad (2)$$

where $\Delta_B = \frac{1}{2}(\sqrt{\varepsilon^2 + 4(\hbar t_c)^2}) + \varepsilon)$ and $\gamma_s$ is the effective tunneling rate for spin-flip inelastic cotunneling ($\gamma_s \equiv \gamma_L = \gamma_R$ for a symmetric barrier is assumed in order to simplify the expression). Eq. (2) can only be applied at positive $\varepsilon$ because the Coulomb blockade dominates at $\varepsilon < 0$.

A current peak with NDC appears when the linear current at the onset of the spin blockade is higher than the blockade current ($G_0V_{peak} > I_{SB}$). From Eqs. (1) and (2), the peak is expected at $\gamma_e$ > $\gamma_s$. Second-order cotunneling with a strong asymmetric barrier satisfies this condition. In addition, the higher-order tunneling process would strongly enhance the elastic cotunneling rate $\gamma_e$ and make $\gamma_e$ >> $\gamma_s$. When the current peak is well defined, we can relate $V_{peak}$ to $\Delta_B$, such that

$$eV_{peak} = \Delta_B = \frac{1}{2}(\sqrt{\varepsilon^2 + 4(\hbar t_c)^2} + \varepsilon). \quad (3)$$

The solid curves in Figs. 4(a-f) are calculated from Eqs. (1-3) with parameters $\hbar t_c = 2\mu eV$, $\hbar\gamma_e = 16.2\mu eV$, and $\hbar\gamma_s = 9.3\mu eV$. The fitting agrees well with the data in Figs. 4(a), (b), and (d) but deviates from those in Figs. 4(c), (e), and (f). In addition, the fitting parameter $\hbar t_c = 2\mu eV$ near the thermal energy $k_BT_e$ (~ 2.2 μeV; $k_B$ is Boltzmann's constant), which does not satisfy our assumption. Thus, this model does not explain our data well.

Since our measurement was performed with relatively weak interdot coupling $t_c$, the current peak might be induced by the thermal effect. Therefore, in the second model, we discuss the peak by considering the thermal effect but completely neglecting coherent coupling (an extreme case with zero ST splitting). As shown in Fig. 3(d), the degenerated (1,1) singlet (S) and triplet (T) states cross the (0,2) singlet state (D) at ε = 0. Transitions between these states are shown by the arrows in the inset. The cotunneling rate between the S (or T) and D states, $\Gamma_\pm$, for positive and negative current contributions and that between the S and T states, $\Gamma_{st}$, are obtained by considering second-order tunneling processes. Inelastic interdot tunneling rate $W_\pm$ depends on ε with $T_e$ = 25 mK. Cotunneling current $I_{cot}$ is obtained by solving the rate equations for these transitions. The assumption of $\Gamma_{st} << \Gamma_\pm << W_\pm$ ensures the appearance of a clear peak with NDC. $G_0$ and $I_{SB}$ are given by

$$G_0 = \left.\frac{dI_{\cot}}{dV_{SD}}\right|_{V_{SD}=0} = \frac{e^2}{4h}\hbar^2\gamma_e^2 V^{-2} \frac{12\varepsilon}{kT(1 + e^{\varepsilon/k_BT_e} - 2e^{-\varepsilon/k_BT_e})} \quad (4)$$

$$I_{SB} = I_{\cot}\big|_{eV_{SD}>>\varepsilon} \approx e\Gamma_{ST} = \frac{e}{h}(\hbar\gamma_s)^2 V^{-2} k_B T_e, \quad (5)$$

which are independent of $W_\pm$ under the assumption. $V_{peak}$ corresponds to the $V_{SD}$ value satisfying $\frac{dI_{\cot}}{dV_{SD}} = 0$ . Dashed lines in Figs. 4(a-f) were obtained from the above formula with $\hbar\gamma_e = 7.9\mu eV$ , $\hbar\gamma_s = 6.2\mu eV$ , and $k_B T_e = 2.2\mu eV$ . It is clear that this simple model explains the data well.

In summary, we analyzed spin blockade transport through a DQD with more than two electrons. Although the peak with NDC might be driven by the thermal effect in our case, we believe that our discussion can be applied to study ST splitting when the interdot coupling $t_c$ is increased. Systematic analyses by the two models should reveal transitions from incoherent tunneling to coherent tunneling in connection with the ST splitting. Also we can extend our work on other effective two-electron DQDs designed by filling Landau-levels[16] and shell structures.[17] Investigating spin correlations in such systems is an important subject for many-body physics as well as controlling spin correlations.

We acknowledge K. Ono and Y. Tokura for useful discussions. This work was partly supported by the Strategic Information and Communications R&D Promotion Program (SCOPE) of the Ministry of Internal Affairs and Communications of Japan, and by a Grant-in-Aid for Scientific Research from the Japan Society for the Promotion of Science.

**87**, 066801 (2001).

**FIG**. **1**. (color online) Current $I$ as a function of gate voltages $V_{GL}$ and $V_{GR}$ at $V_{SD}$ = 110 μV. Hexagonal domains are accentuated by solid lines, where charge states (n,m) denote the effective electron numbers in dots L (n) and R (m). Inset shows an SEM image of our double quantum dot device.

**FIG**. **2**. (color online) (a) Schematic charge diagram in the $V_{GL}$-$V_{GR}$ plane. Sequential and charge-conserved cotunneling regions are shown by hatched and dotted patterns, respectively. (b-e) Absolute current $|I|$ in the $V_{GL}$-$V_{GR}$ plane in regions I (b and c) and II (d and e) of Fig. 1 at positive and negative $V_{SD}$. Dashed lines trace the charge domains described in (a). Two sweeping voltages $V_\delta$ and $V_\varepsilon$ are introduced for convenient operation (See text).

**FIG**. **3**. (color) (a) Current $I$ as a function of $V_\varepsilon$ and $V_{SD}$ at $\delta = 0$. Note that the color scale is different for positive and negative current. (b) $I$-$V_{SD}$ curves at some constant ε, extracted from (a). Arrows indicate the peak position $V_{peak}$ (unit: μV). (c) Anti-crossing energy diagram for the two-electron DQD with coherent interdot coupling. Inset shows the cotunneling transitions between the energy levels. (d) Schematic energy diagram of the DQD with incoherent interdot coupling. Inset shows some possible transitions between the two-electron charge states.

**FIG**. **4**. Zero-bias conductance $G_0$ (a and b), spin blockade current $I_{SB}$ (c and d) at $V_{SD}$ = 50 μV and current peak position $V_{peak}$ (e and f) as a function of ε and δ. Note that (a), (c) and (e) are obtained at $\delta = 0$ and that (b), (d) and (f) are for $\varepsilon = 0$. Solid and dashed curves were calculated considering coherent and incoherent interdot coupling, respectively. Dash-dotted Lorentzian curves in (b) are guides for the sequential tunneling regions and the electrostatic coupling energy U = 255 μeV is determined from the energy spacing between peaks A and B. The dotted line in (e) is a reference for $eV_{peak} = \varepsilon$. The cotunneling region is confined to the δ range of -90 to 90 μeV (between the two solid straight lines).

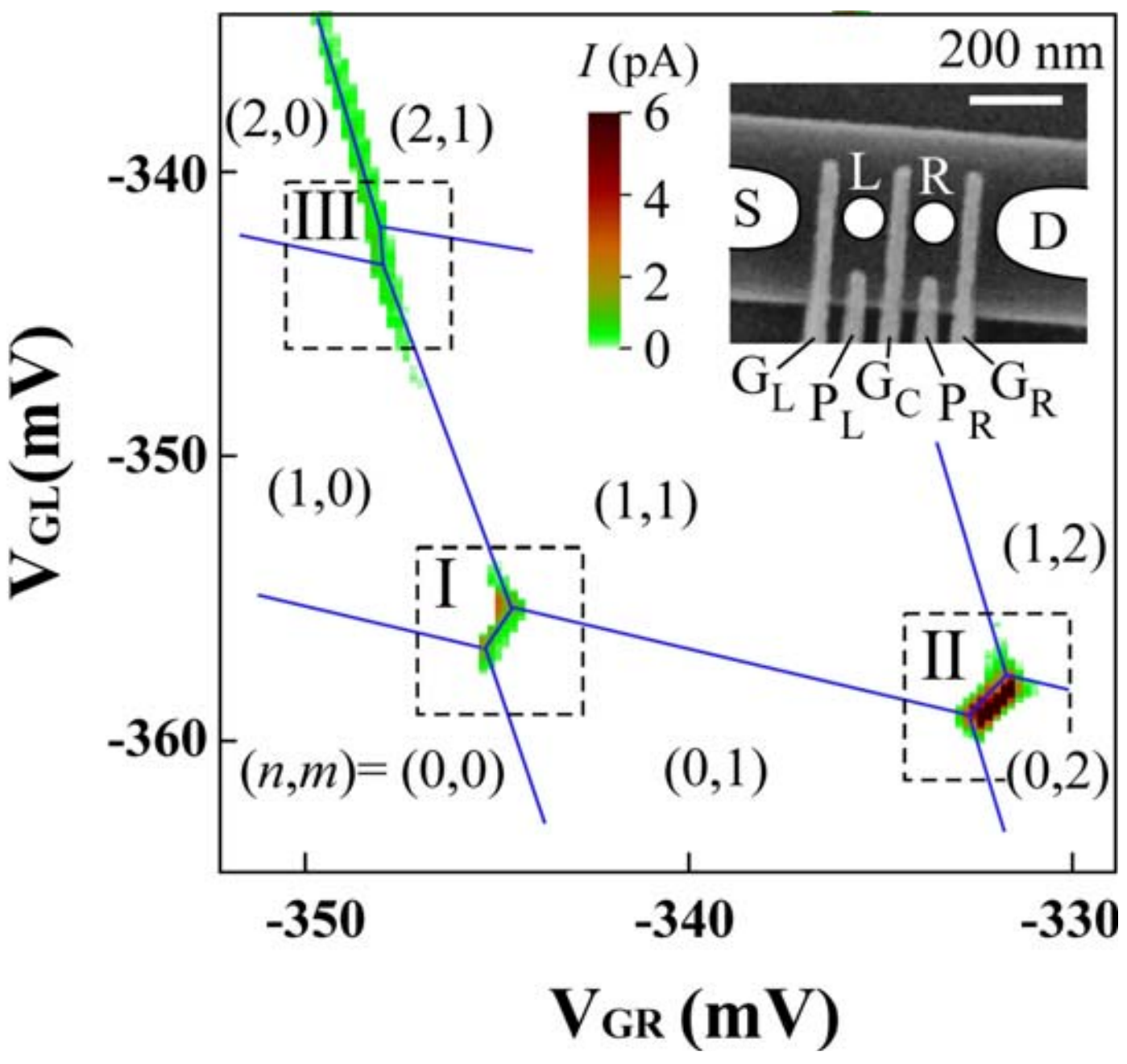

Figure 1

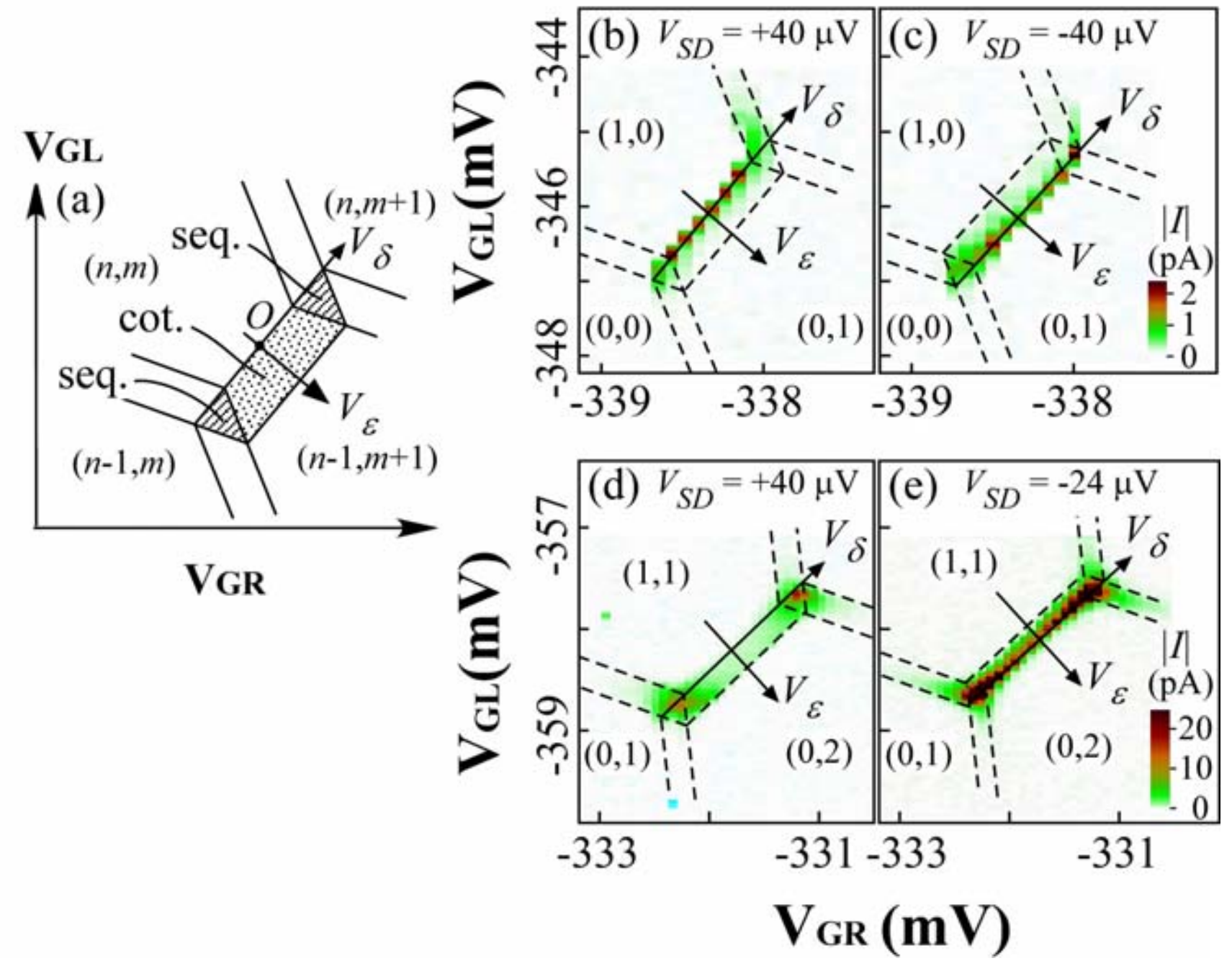

Figure 2

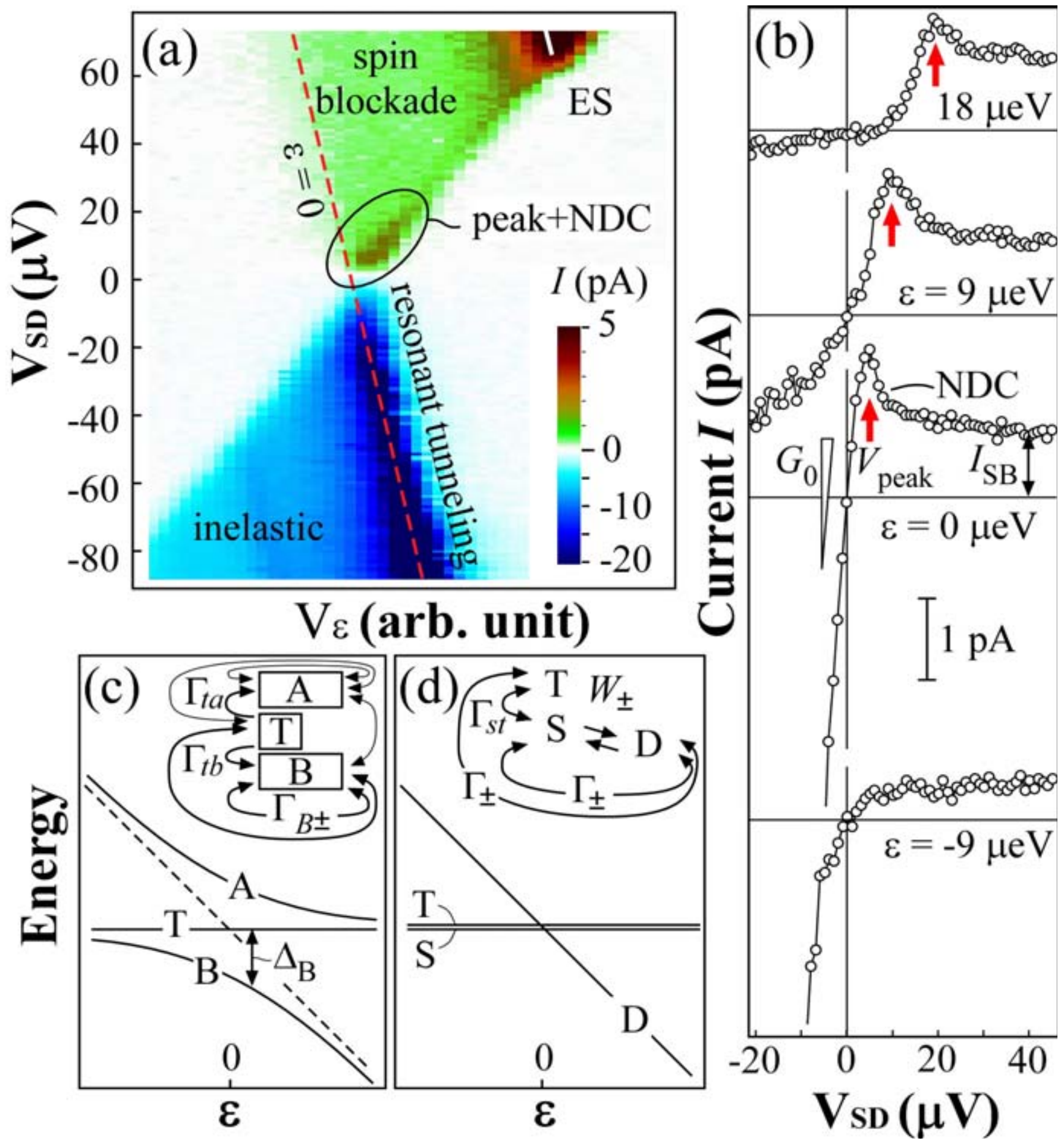

Figure 3

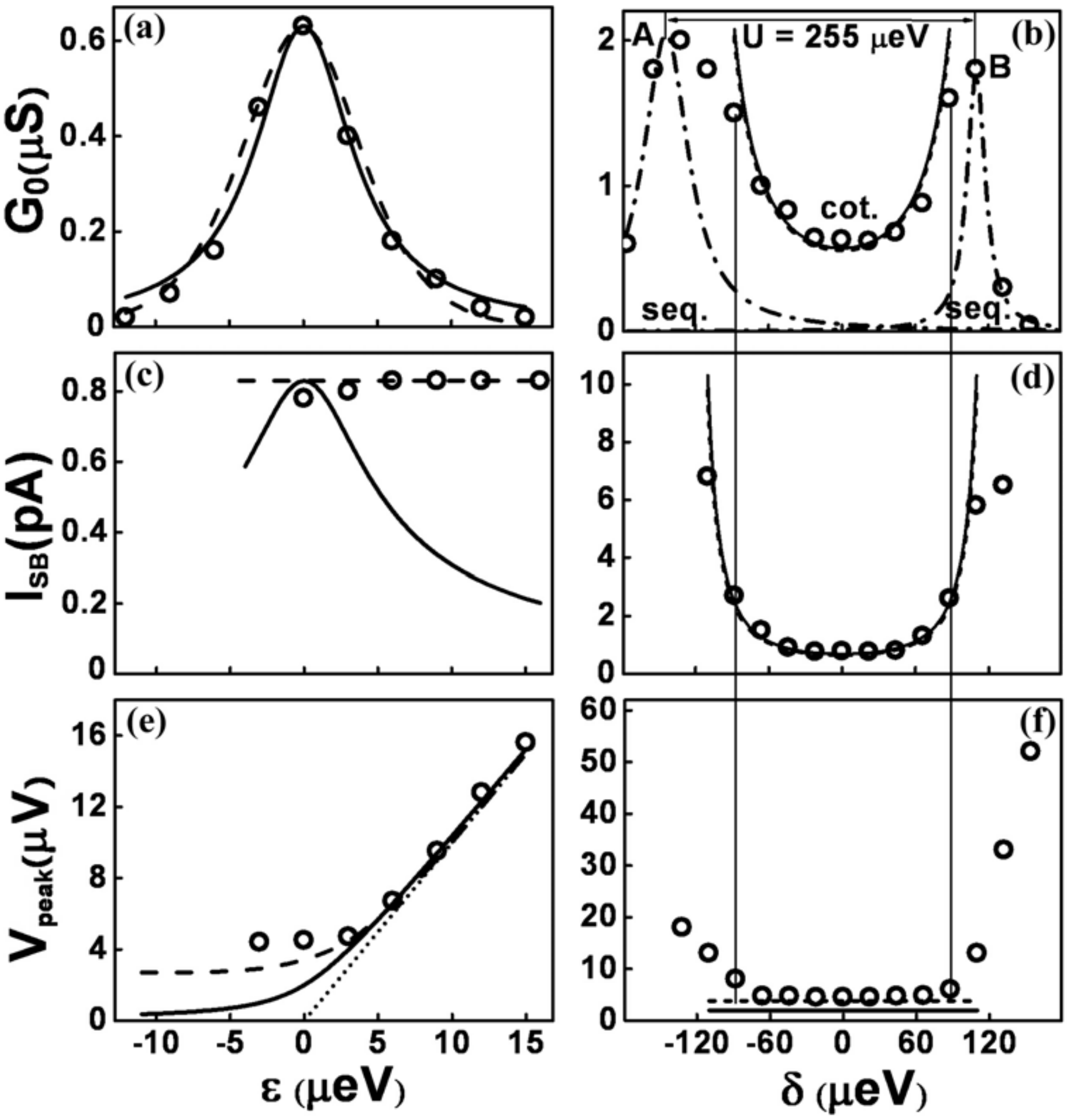

Figure 4